\documentclass[12pt]{article}
\usepackage{graphicx}

\begin{document}

\title{\bf{Observation of An Evolving Magnetic Flux Rope Prior To and During A Solar Eruption}}

\maketitle
{\Large  \bigskip{}
 \bigskip{}
 }{\Large \par}

\author{Jie Zhang$^{1,\ast}$, Xin Cheng$^{1,2}$, Ming-de Ding$^{2}$ \\
\\
\normalsize{$^{1}$School of Physics, Astronomy and Computational Sciences,}\\
\normalsize{George Mason University, Fairfax, VA 22030, USA}\\
\normalsize{$^{2}$Department of Astronomy, Nanjing University, Nanjing 210093, China}\\
\normalsize{$^\ast$To whom correspondence should be addressed; E-mail: jzhang7@gmu.edu.}
}

\newpage{}


\setcounter{page}{1}

\textbf{
Explosive energy release is a common phenomenon occurring in magnetized plasma
systems ranging from laboratories, Earth's magnetosphere,
the solar corona and astrophysical environments.
Its physical explanation is usually attributed to
magnetic reconnection in a thin current sheet.
Here we report the important role of magnetic flux rope structure,
a volumetric current channel, in producing explosive events.
The flux rope is observed as a hot channel prior to and during a solar eruption
from the Atmospheric Imaging Assembly (AIA) telescope
on board the Solar Dynamic Observatory (SDO).
It initially appears as a twisted and writhed sigmoidal structure with a temperature
as high as 10 MK and then transforms toward a semi-circular shape
during a slow rise phase, which is followed by fast
acceleration and onset of a flare.
The observations suggest that the instability of the magnetic
flux rope trigger the eruption, thus making a major addition to
the traditional magnetic-reconnection paradigm.
\vspace{0.5cm}}


In contrast to a thin current sheet structure,
a magnetic flux rope is a volumetric current channel with helical magnetic field
lines wrapping around its center axial field.
Both structures can store a large amount of free magnetic energy in
the current-carrying magnetic fields.
While it is generally believed that magnetic reconnection occurring in the
current sheet releases magnetic energy producing various explosive
phenomena~\cite{parker57, sweet58, yamada99, priest00, kulsrud05},
the role of magnetic flux ropes in the explosive process is
less studied and not well understood.
Nevertheless, there is recently an increasing interest in flux ropes, especially
when a realistic 3-D setting is considered.
Laboratory experiments show that two parallel flux ropes
can drive magnetic reconnection
through magnetohydrodynamic (MHD) attraction~\cite{intrator09}.
Large scale 3-D numerical simulations have demonstrated that
the magnetic reconnection layer is dominated
by the formation and interaction of flux ropes~\cite{daughton11}.
It has been suggested that the observed episodic ejection of plasma blobs in many
black hole systems is caused by the formation and ejection of flux ropes~\cite{yuan09}.

In the research area of solar and heliospheric physics,
the magnetic flux rope is considered to be a fundamental
structure underlying the phenomenon of coronal mass ejections (CMEs),
a well-known driver of space weather that
may affect critical technological systems in space and on the ground.
The first direct observational evidence of the presence of
an isolated magnetic flux rope in the sun-Earth system
is from near-Earth in-situ solar wind observations of
so-called magnetic clouds~\cite{burlaga81,lepping90}.
Improved coronagraphic observations of CMEs from the Solar and Heliospheric
Observatory (SOHO)
showed that CMEs in the outer corona often contain a circular intensity pattern,
suggesting the presence of a magnetic flux rope~\cite{dere99}.
However, the detection of magnetic flux ropes in the lower corona prior to the
CME formation has been elusive.
In fact, one outstanding controversial issue in solar physics, both
observationally and theoretically, has been when, where and how the magnetic flux rope
is formed.
The phenomenon of solar filaments has been interpreted
as being due to magnetic flux ropes~\cite{rust94,kumar11},
as well as the sigmoidal structures (either forward or reverse S-shaped)
often seen in soft X-ray coronal images~\cite{titov99,mckenzie08,aulanier10}.
The origin of a pre-existing flux rope has been suggested to
come directly from sub-photosphere
emerging into the corona~\cite{gibson02},
or alternatively from a sheared arcade in the corona
through a flux-cancellation process~\cite{green09}.
However, other observers argue that the sigmoid geometry is of
sheared field lines, instead of a flux rope,
prior to an eruption~\cite{moore01}.
In the sheared-arcade scenario, numerical simulations show that
the magnetic reconnection in the corona could transform the sheared
arcade into a flux rope~\cite{antiochos99}.

Here we report on unambiguous observational evidence of the presence of a flux rope
prior to and during a solar eruption on 2011 March 8. The observation was made by
the Atmospheric Imaging Assembly (AIA) telescope~\cite{lemen11} on board
the Solar Dynamic Observatory (SDO).
AIA's unprecedented temporal cadence
of 12 seconds, coupled with multi-temperature coronal EUV passbands,
enables for the first time the clear observations of  the detailed evolution of
eruptive structures in the lower
corona~\cite{liu10, cheng11, schrijver11, Zharkov11}.
The flux rope initially appears as a twisted and writhed sigmoidal structure
with a temperature as high as 10 MK. This hot channel then transforms toward
a semi-circular shape during a slow rise phase. This phase is followed
by the fast acceleration of the hot channel and the onset of an accompanying
flare. These observations suggest that the macroscopic
instability of the flux rope trigger the eruption, thus making a
major addition to the traditional magnetic-reconnection paradigm.

\bigskip{}

\noindent{\bf Results}

\begin{figure}[!ht]
\centering
\includegraphics[width=1.00\linewidth]{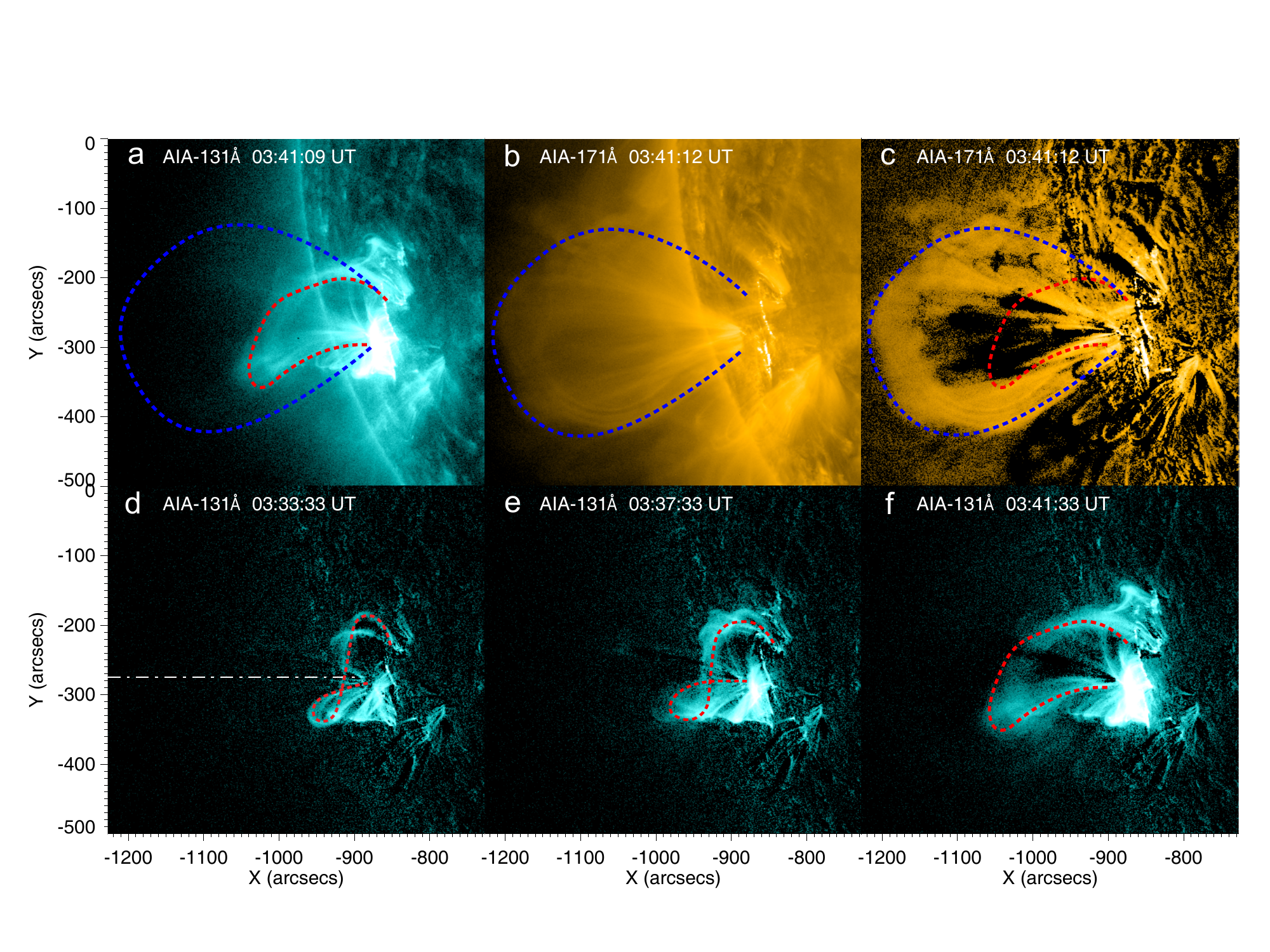}
\caption{{\bf Magnetic flux rope seen as a hot channel in SDO/AIA images.}
The images show three features of the solar eruption on 2011 March 8:
a hot channel (indicated by red dotted lines), a cool compression front
(indicated by blue dotted lines), and a dark cavity.
The time when the image was taken is shown
at the top of each panel.
{\bf (a)} Hot coronal image (131~{\AA}, $\sim$10 MK) showing the hot channel at 03:41:09 UT.
{\bf (b)} Cool coronal image (171~{\AA},$\sim$0.6 MK) at nearly the same time showing
the complete absence of the hot channel.
{\bf (c)} The difference coronal image (171~{\AA}, image at 03:41:12 UT subtracting the
base image at 03:20:41 UT) clearly showing the compression front of the eruption.
It also shows a dark cavity (the center dark region in the image)
forming inside the enveloping compression front.
{\bf (d--f)} A sequence of base-difference images (131~{\AA}, base image at 03:20:09 UT)
showing the evolution of the hot channel. The hot channel apparently transformed
from a writhed sigmoidal shape into a semi-circular shape.
The white dot-dashed line in panel d indicates the
position of a slice at which a slice-time plot is constructed to illustrate
the full evolution of the eruption features.
The full AIA image sequences of the eruption in all six coronal passbands
are provided in the supporting documents (Supplementary Movie 1
and Supplementary Movie 2).
}\label{f1}
\end{figure}

\noindent \textbf{Hot Channel}.
The eruption originated from NOAA Active Region 11171 located at the
heliographic coordinates $\sim$S21E72.
The solar eruption produced a M1.5 class
soft X-ray flare on GOES scale and a CME with a terminal speed of $\sim$700 km/s.
The structural evolution of the eruption is shown in
Fig.~\ref{f1} and Supplementary Movie 1 and 2.
The earliest signature of the hot channel started to appear at
$\sim$03:31 UT (Universal Time),
about six minutes prior to the onset of the flare.
The hot channel only appeared in the AIA's  two hottest
coronal passbands ($\sim$10 MK at 131~\AA\ and $\sim$6.4 MK at 94~\AA),
but completely absent in the cooler passbands
($\sim$1.6 MK at 193~\AA\ and $\sim$0.6 MK at 171~\AA).
The hot channel showed an interesting morphological evolution
as it transformed itself from a sigmoidal shape (Fig.~\ref{f1}d,e)
to a loop-like shape (Fig.~\ref{f1}f).
The initial sigmoid had a twisted or writhed axis,
with its two elbows close to the footpoints
extending into the corona and the center part dipping toward the surface.
The dipped center part then rose up becoming more linear.
The continuing rise of the center part eventually turned the channel
into a loop-like shape, or partial torus.
During this transformation process, the two footpoints of the
evolving hot channel remained fixed.

\begin{figure}[!ht]
\centering
\includegraphics[width=0.90\linewidth]{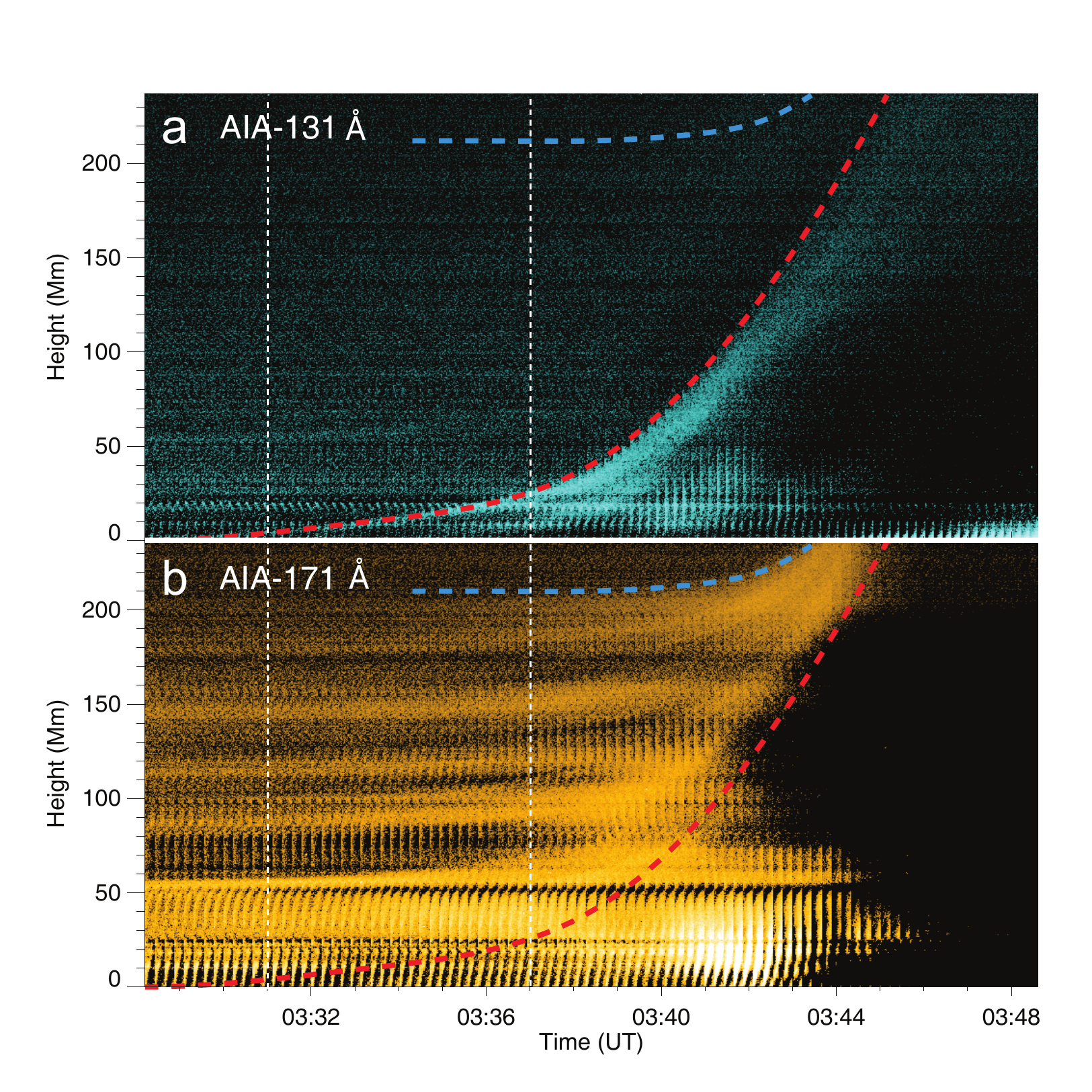}
\caption{{\bf Time evolution of the hot channel and compression front.}
The constructed slice-time plots illustrate the rising motion of the
eruption on 2011 March 8.
{\bf (a)} Slice-time plot of AIA 131~{\AA} images showing the rise motion
of the hot channel. {\bf (b)}slice time plot of 171~{\AA} images showing the
rising motion of the cool compression front.
The positions of the hot channel and the compression front are
outlined by the red and blue dashed lines, respectively.
The white vertical dashed line on the left indicates the start time
of the slow-rise phase of the hot channel. The white vertical dashed
line on the right indicates the onset time of the accompanying solar flare;
apparently, this same line also marks the onset of the fast acceleration
phase of the hot channel.
}\label{f2}
\end{figure}

\begin{figure}[!ht]
\centering
\includegraphics[width=0.90\linewidth]{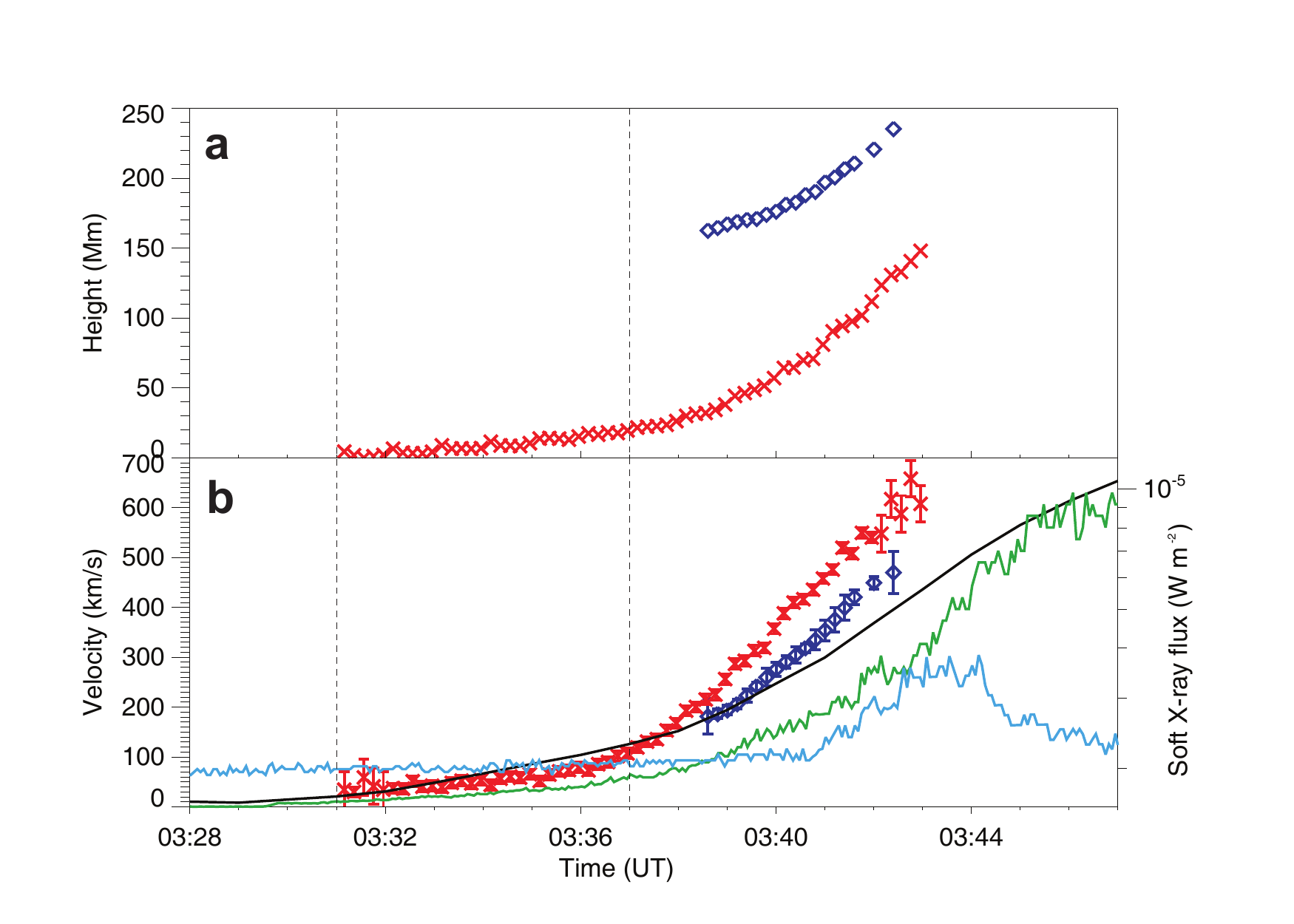}
\caption{{\bf Kinematic evolution of eruption features on 2011 March 8.}
{\bf (a)} the height-time plots of the hot channel (red cross symbols)
and the compression front (blue diamond symbols). The uncertainty
of the height measurement is about 2 Mm, whose error bar size
is much smaller than the symbol sizes.
{\bf (b)} the velocity-time plots of the hot channel
(red cross symbol) and the compression front (blue diamond symbol).
The velocity uncertainty is about 30 km/s. The error bar size
is similar to the symbol size.
The flux profiles of the accompanying flare are overlaid on the velocity
plots: GOES soft X-ray 1--8~{\AA} (black line),
RHESSI hard X-ray 6-12 keV (green line) and 25-50 keV (cyan line).
The vertical dashed line on the left indicates the start time
of the slow-rise phase of the hot channel. The vertical dashed
line on the right indicates the onset time of the accompanying solar flare;
apparently, this same line also marks the onset of the fast acceleration
phase of the hot channel.
}
\label{f3}
\end{figure}

We also carried out a kinematic analysis of the hot channel~(Fig.~\ref{f2}, \ref{f3}).
The entire eruption process as seen by AIA can be divided into
two distinct phases: a slow rise phase
prior to the flare onset and a fast acceleration phase after the flare onset.
The slow rise phase lasted for about six minutes from
$\sim$03:31 UT to 03:37 UT.
The hot channel started to appear in the beginning of this phase and
then slowly and steadily rose up with an average velocity of $\sim$60 km/s;
at the end of this phase, the velocity increased to about $\sim$100 km/s.
This slow rise phase was then followed by a much more energetic phase
that was accompanied by multiple explosive signatures , including
the fast acceleration of the hot channel,
the formation of a cool compression front running ahead of the hot channel,
the growth of a dark cavity, and a flare of electromagnetic radiation.
It has been known that the onset of the fast acceleration phase of CMEs coincides well
with the onset of accompanying flares~\cite{zhang01}.
However, the high cadence observations for this event show that the difference
of the two onset times may be as small as one minute (as indicated
by the vertical line on the right in Fig.~\ref{f2}, \ref{f3}).
During the fast acceleration phase, the velocity
of the hot channel increased from 100 km/s to 700 km/s, with an average
acceleration about $\sim$1600 m/s$^2$, which is at least ten times stronger than
that in the slow rise phase.

The bright but cool front of the eruption, best seen in AIA 171~\AA\
(Fig.~\ref{f1}c, Fig.~\ref{f2}b),
is believed to be a compression front.
It was apparently caused by the compression of the plasma surrounding
the hot channel.
The compression front only formed late in the fast acceleration phase.
The magnetic structure of the active region presumably had two
components: one internal core region composed of a flux rope, and one
external enveloping region consisting of near-potential magnetic fields.
The expansion of the envelope fields formed the bright front through compression.
The expansion also explains the development of a dark cavity behind the compression
front, as best seen in Fig~\ref{f1}c, because of the rarefaction of the volume.
The velocity of the compression front is always smaller than
that of the hot channel throughout the fast acceleration phase~(Fig.~\ref{f3}b),
indicating that the eruption is mainly driven by the hot channel.

\bigskip{}
\noindent \textbf{Flux rope}.
We conclude that the observed evolving hot channel is a
magnetic flux rope.
It is the coherent helical magnetic field lines that maintain
the structure of the channel during the violent eruption
process, even though its center axis changes
from a twisted sigmoidal shape to a semi-circular shape,
and even though the structure undergoes an acceleration as strong as almost six
times the solar surface gravitational acceleration (274 ${\rm m/s^2}$).
The two ends of the flux rope remain anchored onto the photosphere because of the
line-tying effect.
The observed morphological evolution resembles the result of the 3-D numerical
simulation of a magnetic flux rope~\cite{aulanier10}.

We further conclude that the magnetic flux rope
has fully formed before the onset of the eruption. The onset of the eruption
is clearly marked by the onset of the flare and the
fast acceleration of the flux rope.
The initial sigmoidal-shaped flux rope formed even before the slow rise phase.
Therefore, the flux rope should have formed during the period of the
relative quiescent evolution of the source active region,
which can last for days or weeks. It could also form through an interior
dynamo process prior to its emergence to the surface.

\bigskip{}
\noindent{\bf Discussion}

The observational result described above has serious theoretical implications.
It makes a major addition to the standard paradigm of eruptive flares,
the so-called CSHKP model~~\cite{carmichael64,sturrock68,hirayama74,kopp76}.
The CSHKP model and its many variants assume that the current-sheet magnetic
configuration is of primary importance;  the magnetic reconnection in the current
sheet releases the magnetic energy that produces flare emission,
flare ribbons and post-flare loop arcades.
Nevertheless, the standard model does not explicitly address the trigger of
the eruption. Our work provides strong evidence that the trigger
is the instability of a pre-existing flux rope.
The reconnecting current sheet is likely formed underneath the rising flux
rope through the upward stretching of surrounding magnetic fields.
One such mechanism of triggering is the so-called
Torus Instability (TI)~\cite{kliem06,olmedo10},
an ideal MHD process responsible for the loss of equilibrium of a
toroidal current ring. In the TI model, a critical
vertical gradient of the external magnetic field determines the instability.
The appearance of the instability should mark the separation between the
slow rise phase and the fast acceleration phase,
corresponding to the onset of the eruption.
The fast-rising motion of the flux rope may create a current sheet underneath and
drive plasma in-flow toward the current sheet,
resulting in fast magnetic reconnection.
In this scenario, the occurrence of fast magnetic reconnection
is not spontaneous; instead, it is driven by the macroscopic motion of a magnetic
flux rope.
Our result suggests magnetic flux ropes play an important role in triggering
and driving the explosive energy release process in solar eruptions,
and possibly in many other plasma systems in space and laboratories.

\bigskip{}
\noindent{\bf Methods}

\noindent \textbf{Data and Observations}.
The AIA instrument on SDO provides the essential observations of the flux rope.
The AIA instrument has ten passbands, six of which are sensitive to coronal temperatures
mainly contributed from emissions of specific spectral lines.
The six coronal passbands, in the order of decreasing temperature,
are 131~{\AA} (Fe XXI, $\sim$10 MK), 94~{\AA} (Fe XVIII, $\sim$6.4 MK),
335~{\AA} (Fe XVI, $\sim$2.5 MK), 211~{\AA} (Fe XIV, $\sim$2.0 MK),
193~{\AA} (Fe XII, $\sim$1.6 MK),
and 171~{\AA} (Fe IX, $\sim$0.6 MK), respectively.
Each AIA image has 4096~$\times$ 4096 pixels
(0.6$''$ pixel size, 1.5$''$ spatial resolution)
covering the full disk of the Sun and up to 0.5 $R_\odot$ above the limb.
The observational cadence of AIA is one image every 12 seconds
at each passband. The AIA images shown in this paper
(Fig.~\ref{f1} and Supplementary Movies)
are a small portion of the original full size images.

The X-ray data of the flare, as shown in Fig.~\ref{f3}, are from two instruments.
The soft X-ray data are from the Geostationary Operational Environment
Satellite (GOES) that provides the integrated
full-disk soft X-ray emission from the Sun. GOES soft X-ray data have been
historically used to characterize the magnitude, onset time, and peak time of solar
flares. The hard X-ray data are from the Reuven Ramaty High Energy Solar
Spectroscopic Imager (RHESSI) spacecraft.

\noindent \textbf{Kinematic Analysis}.
The kinematic information of the hot channel and the compression front,
as shown in Fig.~\ref{f3}, is obtained by analyzing AIA images.
We visually inspect the images and identify the leading edges of these
features. The heights are measured from
the projected distance of the leading edges from the initial
position of the hot channel. The measurement of the hot channel is
made on AIA 131~\AA\ images, while that of the compression front is on
AIA 171~\AA\ images. The uncertainty of the height measurement is about
4 pixels, or 2 Mm, which is much smaller than the size of the height
symbols used in Fig.~\ref{f3}. Based on the height-time measurements,
the velocities are then derived from
the numerical derivative method that uses Lagrangian
interpolation of three neighboring points. In order to reduce
the uncertainty of the derived velocities, the height points are
smoothed using a cubic spline interpolation method.
The derived velocity uncertainty is about
30 km/s. As shown in Fig.~\ref{f3},
the velocity error bars are about the same size as the symbols, except
for the edge points.

\bibliographystyle{}

\noindent{\bf Acknowledgments}

SDO is a mission of NASA's Living With a Star Program. J.Z. is supported by NSF grant
ATM-0748003 and NASA grant NNX07AO72G. X.C. and M.D. are supported by NSFC under
grants 10828306, and 10933003 and NKBRSF under grant 2011CB811402.
X.C. is also supported by the scholarship granted by the China Scholarship
Council (CSC) under file No. 2010619071.

\bigskip{}

\noindent{\bf Author contribution}\\
{
J. Z. developed the idea and led the discussion.
X. C. carried out the data analysis.
M. D. contributed to discussion. All wrote the paper.
}

\bigskip{}

\noindent {\bf Competing financial interests:}
{The authors declare no competing financial interests.}

\bigskip{}

\noindent {\bf Additional information}\\
{Correspondence and requests for materials should be address to J. Z.}

\bigskip{}

\noindent {\bf Supporting Material: }\\
There are two pieces: Supplementary Movie 1 and supplementary Movie 2.

\newpage

\end{document}